# Scheme for suppressing atom expansion induced contrast loss in atom interferometers

Qing-Qing Hu[1,2], Yu-Kun Luo[1,2], Ai-Ai Jia[1,2], Chun-Hua Wei[1,2], Shu-Hua Yan[1,2], Jun Yang[1,2†]

[1]*College of Mechatronic Engineering and Automation, National University of Defense Technology, Changsha, Hunan 410073, China*

[2]*Interdisciplinary Center for Quantum Information, National University of Defense Technology, Changsha, Hunan, 410073, China*

**Abstract**

The loss of contrast due to atom expansion induced non-perfect Raman pulse area in atom interferometers is investigated systematically. Based on the theoretical simulation, we find that the expansion of the atomic cloud results in a decrease of the π pulse fidelity and a change of the π pulse duration, which lead to a significant reduction in fringe contrast. We propose a mitigation strategy of increasing the intensities of the second and third Raman pulses. Simulation results show that the fringe contrast can be improved by 13.6% in a typical atom interferometer gravimeter using this intensity compensation strategy. We also evaluate the effects of this mitigation strategy in the case of a lower atomic cloud temperature and a larger Raman beam size under different Raman pulse time interval conditions. This mitigation strategy has potential applications in increasing the sensitivity of atom interferometer-based precision measuring, including precision measuring of the gravity, gravity gradient, rotation, and magnetic field gradient, as well as testing of the Einstein equivalence principle.

## 1. Introduction

Since the first light pulse cold atom interferometer was achieved by Steven Chu's team in 1991[1], atom interferometers (AIs) have been demonstrated remarkable prospects. Applications of AIs stretch from high-precision measurement of local gravity[2], gravity gradient[3, 4] and Sagnac effect[5], to the measurement of physical constants, such as fine structure constant[6], gravitational constant[7], quadratic Zeeman coefficient[8] and so on. Improving the measurement sensitivity of the AIs further is one of the main tasks in fundamental research and practical application.

† Corresponding author. E-mail: jyang@nudt.edu.cn

The single-shot acceleration measurement sensitivity of a quantum projection noise (QPN) limited[9, 10] Mach-Zehnder (M-Z) AI with $N$ atoms can be expressed as $S \propto 1/(k_{eff}T^2C\sqrt{N})$. Obviously, the sensitivity will benefit from enlarging the transferred momentum associated $k_{eff}$, the interval time $T$, the contrast $C$, and the atom number $N$. Therefore, the large momentum transfer (LMT) schemes including sequential two-photon Raman transition[11, 12], $n$-order Bragg diffraction[13, 14], Bloch oscillations in optical lattices[15, 16], the long baseline schemes including AIs of 10 m[17], 120 m[18] and AIs in space microgravity environment[19], as well as the high-flux cold atom source schemes[20] have been proposed in order to increase the measurement sensitivity of AIs. Recently, Kasevich's group[21] achieved the best acceleration sensitivity of $6.7\times10^{-12} g$ in one shot by enlarging the interval time $T$ to 1.15 s with the Bose-Einstein-condensation (BEC) atom source of 50 nK temperature. However, the evaporation cooling setups used to create the 50 nK BEC are too bulky and complicated to be included in a mobile atom gravimeter[22-24] for practical application. Second only to Kasevich's group, Hu et al.[25] achieved a short-term sensitivity of $4.2\times10^{-9}\ g/\sqrt{Hz}$ with a shorter interval time of 300 ms and a larger atom source temperature of 7 $\mu$K using Magneto-Optical Trap (MOT). Although a velocity selection pulse[26, 27] has been implemented to suppress atom vertical velocity distribution, the influence due to atom horizontal expansion still exists and leads to a non-perfect Raman pulse area and a decay in fringe contrast. As a result, the contrast in Hu's atom gravimeter is only 16%[25].

From the sensitivity function above, it is obvious that the sensitivity of the atom gravimeter can be increased further by improving the fringe contrast. There are a number of factors lead to contrast reduction, the most significant one is that the atoms still have a rather broad velocity distribution perpendicular to the Raman beam axis. This causes a number of atoms to move away from the center of the Gaussian Raman beam during the flight, where the light intensity is lower. Consequently, for these atoms a pulse length of $2\tau$ does not amount to a $\pi$ pulse anymore, which causes many atoms to be in the wrong state at the output port of the interferometer. The effects of the Gaussian distribution of laser intensity and atomic cloud in an atom gravimeter have been investigated from aspects of the light-shift-related uncertainty[28], the amplitude and phase shift of the interference signal[29], and the bias caused in atom gravimeter[30]. However, there is no article discussing the corresponding mitigation strategy for contrast reduction. In contrast, other factors, including gravity gradients[31], alignment defects[32], vibrations and imperfect separation of the exit beams[33], related contrast reduction and the corresponding mitigation strategies have been studied already. In this paper, we focus on the mechanism of contrast reduction due to atom expansion, building a mathematical model to analyze the effects of atom expansion on Raman $\pi$ pulse fidelity and on fringe contrast, putting forward a simple and easy implemented mitigation strategy, and evaluating the effects of this mitigation strategy in typical atom interferometer gravimeter conditions and in different Raman pulse time interval conditions.

## 2. Theoretical model

The basic principle of a Mach-Zehnder type atom interferometer is shown in Fig. 1. It is realized by a $\pi/2$–$\pi$-$\pi/2$ Raman pulse sequence to coherently split, reflect, and recombine the atomic wave packet. In large detuning condition, the wave function of a three-level atom interacting with a Raman pulse can be described by the linear superposition of a two-level atom ground eigenstates

$$|\psi(t)\rangle = c_1(t)|1\rangle + c_2(t)|2\rangle, \text{ with } c_1^2(t) + c_2^2(t) = 1, \tag{1}$$

and the time evolutions of the atom eigenstate coefficients in stimulated Raman transition are given by[27]

$$\begin{pmatrix} c_1(t) \\ c_2(t) \end{pmatrix} = M(\Omega_R, \tau, \delta, t_0, \phi) \begin{pmatrix} c_1(t_0) \\ c_2(t_0) \end{pmatrix}, \tag{2}$$

in which the transfer function

$$M(\Omega_R, \tau, \delta, t_0, \phi) = \begin{bmatrix} A & -iB \\ -iB^* & A^* \end{bmatrix}, \text{ with } \begin{cases} A = \cos(\Omega_R \tau/2) - i\sin\alpha \sin(\Omega_R \tau/2) \\ B = \exp(i(\delta t_0 + \phi))\cos\alpha \sin(\Omega_R \tau/2) \end{cases}, \tag{3}$$

where $\tau$ is the Raman pulse duration, $t_0$ is the Raman pulse irradiation moment, $\phi$ is the Raman laser phase, and $\delta = 2k\sigma_{v_z}$ represents the two-photon Raman detuning caused by atom vertical velocity distribution $\sigma_{v_z}$ (assuming the Doppler frequency shift caused by gravity is compensated completely), in which $k$ is the wavenumber of the Raman laser field. $\alpha$ is determined by

$\sin\alpha = \delta/\Omega_R$, $\cos\alpha = \Omega_{eff}/\Omega_R$, and $\Omega_R = \sqrt{\Omega_{eff}^2 + \delta^2}$ is the Rabi frequency including two-photon detuning,. $\Omega_{eff} = \Omega_{max} \exp\left[-(\frac{r^2}{2w^2})\right]$ is the effective two-photon Rabi frequency, and $\Omega_{max} = \frac{\Gamma^2}{2I_s\delta}\frac{P_0}{\pi w^2}$ is the maximum two-photon Rabi frequency when an atom stays at the center of the Raman beam, in which $\Gamma$ is atom natural line width, $I_s$ is the saturated absorption intensity, $P_0$ and $w$ are the total power and diameter of the Raman beam, respectively.

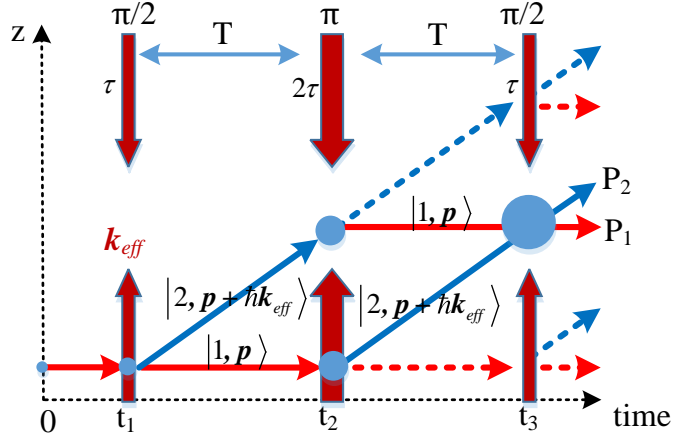

Fig. 1. The principle of a Mach–Zehnder type atom interferometer. The horizontal axis is atom free flight time after launch, and the vertical axis is the relative position of the atom compared to a reference frame that is in free fall along the initial trajectory of the atom. $|1\rangle$ and $|2\rangle$ represent the atom two ground eigenstates, $|p\rangle$ and $|p + \hbar k_{eff}\rangle$ represent the atom initial and finial momentum state during two-photon Raman transition. Dashed lines represent the non-perfect Raman π pulse related atom loss.

Ideally, the atoms move with the same velocity and hence the two-photon Raman detuning $\delta \approx 0$, and the Raman beam intensities are homogeneous and hence all atoms experience the same Raman pulse area $\Omega_R\tau$. In this situation, a perfect splitting (50:50) or mirror (0:100) pulse is achieved when $\Omega_R\tau = \pi/2$ or $\pi$. However, in experiment, we have to consider the velocity distribution of the atomic cloud and the intensity distribution of the Raman beam because of the limited atomic cloud temperature. Usually, the position and velocity distribution of a cold atomic cloud follow a normal Gaussian distribution, and the probability of finding an atom in the phase space volume element with coordinate $(x_0, y_0, z_0; v_{x0}, v_{y0}, v_{z0})$ is given by[34]

$$N(x_0, y_0, z_0; v_{x0}, v_{y0}, v_{z0}) = \prod_{\xi \in \{x,y,z\}} g(\xi_0, \sigma_{\xi_0}) g(v_{\xi 0}, \sigma_{v_{\xi 0}}), \tag{4}$$

where $g(x, \sigma) = (2\pi\sigma^2)^{-1/2} \exp(-x^2/2\sigma^2)$ is the general expression for a normalized one-dimensional Gaussian distribution. We can see that the horizontal position distribution and vertical velocity distribution are independent from each other. Therefore, to simplify the considerations, we focus on atom horizontal expansion induced contrast reduction and ignore the influence of the atom vertical velocity distribution in this article (actually, the vertical velocity distribution would cause two-photon Raman detuning and decrease the fringe contrast again), which means that in the following discussion, we assume $\delta = 0$ in equation (3). The atom density distribution in cylindrical coordinate system can be written as

$$n(r, z, t) = n_0 \exp\left(-\frac{r^2 + z^2}{2\sigma(t)^2}\right), \tag{5}$$

where $n_0$ is the maximum atom density at the center of the Raman beam, and the 1/e Gaussian width $\sigma(t)$ with flying time $t$ after launch is

$$\sigma(t) = \sqrt{\sigma_0^2 + \sigma_{v_r}^2 t^2}, \tag{6}$$

in which $\sigma_0$ is the initial size of the cold atomic cloud, $\sigma_{v_r} = \sqrt{k_B T_{emp}/M}$ is the temperature $T_{emp}$ dependent atom velocity distribution width, $k_B$ is the Boltzmann constant and $M$ is the atom mass.

Assuming all atoms are in the lower eigenstate ($\boldsymbol{C}_0 = \begin{bmatrix} c_1(0) \\ c_2(0) \end{bmatrix} = \begin{bmatrix} 1 \\ 0 \end{bmatrix}$) at initial moment, and combing equation (2), (3) and (5), the atom transition probability from state $|1\rangle$ to state $|2\rangle$ after a Raman pulse can be written as

$$\begin{aligned} P_2 = |c_2|^2 &= \int\int n(r,z,t) \left[ \boldsymbol{M}(\Omega_R, \tau, 0, t_0, \phi) \boldsymbol{C}_0 \right]^2 drdz \\ &= \int_{-\infty}^{\infty}\int_0^{\infty} \frac{r}{\sqrt{2\pi}\,\sigma(t)^3} \exp\left(-\frac{r^2+z^2}{2\sigma(t)^2}\right) \sin^2\left\{ \frac{\tau \cdot \Omega_{\max}}{2} \exp\left[-\left(\frac{r^2}{2w^2}\right)\right]\right\} drdz \end{aligned} \tag{7}$$

Extending the above method to the three pulse M–Z AI case (Fig. 1), the final transition probability at the output port of the AI is

$$P_{2tot} = \int_{-\infty}^{\infty}\int_{0}^{\infty} n(r,z,t)\left[\boldsymbol{M}\left(\Omega_{R_3},\tau_3,0,t_3,\phi_3\right).\boldsymbol{M}\left(\Omega_{R_2},\tau_2,0,t_2,\phi_2\right).\boldsymbol{M}\left(\Omega_{R_1},\tau_1,0,t_1,\phi_1\right)\boldsymbol{C}_0\right]^2 drdz. \quad (8)$$

Deriving the analytic expression of the integrals for equation (7) and (8) is quite difficult, thus we use numerical calculation to analyze the effects of the related parameters. Based on equation (7) and (8), the fidelity of the Raman $\pi$ pulse and the fringe contrast of the AI are

$$F = P_2\big|_{\tau=\pi/\Omega_{max}} \quad (9)$$

$$C = \frac{P_{2tot}\big|_{\phi_3=\pi} - P_{2tot}\big|_{\phi_3=0}}{P_{2tot}\big|_{\phi_3=\pi} + P_{2tot}\big|_{\phi_3=0}} \quad (10)$$

### 3. Contrast loss due to atom expansion
### 3.1 Time variation of the atomic cloud diameter

As shown in Fig. 1, the spatial volume of the atomic cloud is increasing with free flight time because of the thermal expansion. We write the diameter of the atomic cloud at the irradiation moment of the $i^{th}$ Raman pulse as $\sigma_i (i=1,2,3)$. Without loss of generality, we consider $^{87}$Rb atom in the following calculations. According to equation (6), we simulate the time variation of the atomic cloud diameter and the corresponding diameter ratio of Raman beam to atomic cloud $s = w/\sigma$ (shortened as diameter ratio in the below paragraphs) with two sets of parameters. One represents the typical normal situation ($T_{emp} = 7\,\mu K$, $w = 20\,mm$)[25, 35] and the other represents the better situation (lower atom temperature $T'_{emp} = 3\,\mu K$ and larger Raman beam $w' = 30\,mm$)[22, 36]. The initial diameters of atomic cloud in both situations are assumed equal to $\sigma_0 = 3\,mm$. As shown in Fig. 2, after one second of free flight ($t=0$ corresponding to the atom launch moment), the diameter of the atomic cloud increases to 26.1 mm in the normal situation (black triangles) and increases to 17.2 mm in the better situation (black rectangles). The diameter ratio decreases from 6.7 to 0.8 in the normal situation (red triangles) and decreases from 10 to 1.7 in the better situation (red rectangles).

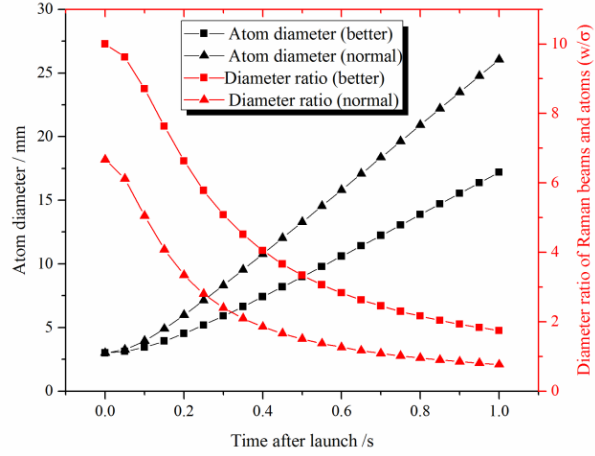

Fig. 2. Time variation of the atomic cloud diameter (black) and the corresponding diameter ratio (red) of Raman beam to atomic cloud under a typical normal situation (triangles, $T_{emp} = 7\,\mu K$, $w = 20\,mm$) and a better situation (rectangles, $T'_{emp} = 3\,\mu K$, $w' = 30\,mm$).

## 3.2 The effect of atom expansion on Raman π pulse fidelity

Considering that the fidelity is determined by the diameter ratio rather than the absolute diameter for either of the atomic cloud or Raman beam (this can be verified by calculating the integrals of equation (7) with some varying but proportional diameters of Raman beam and atomic cloud), we analyze the effects with diameter ratio instead. Adopting the experimental parameters in [36], the irradiation moment of the first Raman pulse in the M-Z interferometer is $t_1 = 130\,ms$, and the time interval between Raman pulses is $T = 260\,ms$, we can obtain the corresponding diameter ratios at the three Raman pulse irradiation moments from Fig. 2 as $s_1, s_2, s_3 = 4.4, 1.9, 1.2$ respectively in the normal situation, and $s'_1, s'_2, s'_3 = 8.1, 4.1, 2.6$ respectively in the better situation. Substituting these diameter ratios into equation (9), setting $t_0 = 0$, $\phi = 0$, $\Omega_{max} = \Omega_{max0}$ for simplicity, and calculating the integrals with a varying Raman pulse duration $\tau$, we can obtain the Rabi oscillation curves as shown in Fig. 3 (a). It shows that the decrease of the diameter ratio results in a decrease of the π pulse fidelity and a change of the π pulse duration. These results can be explained by Fig. 3 (b), in which the expansion of the atomic cloud results in a larger Gaussian distribution width and lets atoms experience a larger Raman laser intensity inhomogeneity and a lower average intensity.

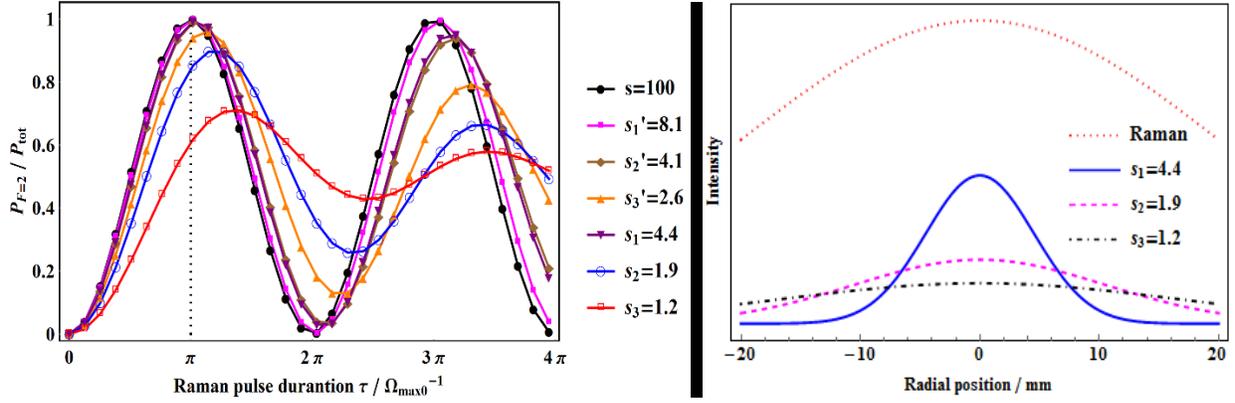

Fig. 3. (a) Rabi oscillation curves under different diameter ratios. It shows that the decrease of the diameter ratio results in a decrease of the $\pi$ pulse fidelity and a change of the $\pi$ pulse duration. (b) Distributions of Raman beam intensity and atom density at the three Raman pulse irradiation moments in the normal situation ($T_{emp} = 7\,\mu K$, $w = 20\,mm$, $s_1, s_2, s_3 = 4.4, 1.9, 1.2$). The expansion of the atomic cloud results in a larger Gaussian distribution width and lets atoms experience a larger Raman laser intensity inhomogeneity and a lower average intensity.

### 3.3 The effect of atom expansion on fringe contrast

Substituting the Raman irradiation moment $t_i$ and the corresponding diameter ratios into equation (8), setting $\Omega_{max} = \Omega_{max0}$, $2\tau_1 = 2\tau_3 = \tau_2 = \pi/\Omega_{max0}$, $\phi_1 = \phi_2 = 0$ for simplicity, and calculating the integrals with a varying phase of $\phi_3$, we can obtain the interferometer fringes in the normal, better, and ideal ($T''_{emp} = 0\,\mu K$, diameter ratio $s''_1 = s''_2 = s''_3 = 100$) situations as shown in Fig. 4. The fringe contrasts in these three situations calculated with equation (10) are 41.5%, 87.4%, and 100%, respectively. We can see a significant fringe contrast reduction as the decrease of the diameter ratio.

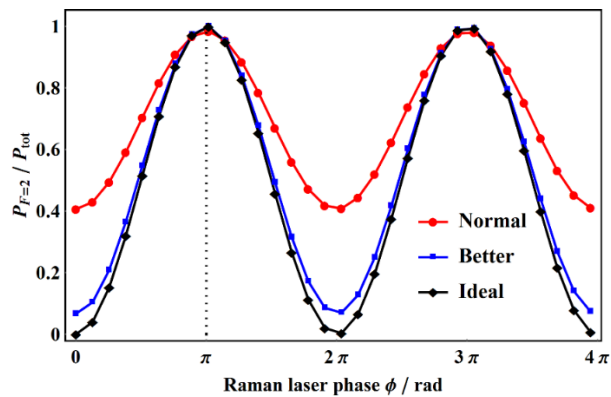

Fig. 4. Interferometer fringes under the normal ($T_{emp} = 7\,\mu K$, $w = 20\,mm$, $s_1, s_2, s_3 = 4.4, 1.9, 1.2$), better ($T'_{emp} = 3\,\mu K$, $w' = 30\,mm$, $s'_1, s'_2, s'_3 = 8.1, 4.1, 2.6$), and ideal ($T''_{emp} = 0\,\mu K$, $s''_1 = s''_2 = s''_3 = 100$) situations. The fringe contrasts in these three situations are 41.5%, 87.4%, and 100% respectively.

## 4. Mitigation strategy

### 4.1 Implement of the mitigation strategy

As shown in Fig. 3 (b), the atomic cloud would experience a larger Raman laser intensity inhomogeneity and a lower average intensity because of expansion. The intensity inhomogeneity is hardly to be compensated because enlarging the diameter of the Raman beam during a measurement cycle of one second range is technically impractical. However, since the laser intensity is linearly proportional to the driving volt of the acoustic-optic modulator (AOM), the average intensities of the Raman beams can be changed by changing the Raman laser AOM attenuator's driving volt in the timing sequence. In experiment, similar to Fig. 3 in reference[37], one can find the optimal duration time $\tau_i$ ($i$=1, 2, 3) of the $i$th Raman pulse in order to achieve a maximum $\pi$ pulse fidelity by achieving the corresponding Rabi oscillation curves of the $i$th Raman pulse. Then, one can achieve a higher $\pi$ pulse fidelity and a higher fringe contrast by increasing the Raman laser AOM attenuator's driving volt accordingly in the timing sequence. That is, to achieve a higher fringe contrast, the only thing one need to do is achieving the Rabi oscillation curves for the $i$th Raman pulse and changing the AOM's driving volt accordingly in the timing sequence.

### 4.2 Raman $\pi$ pulse fidelity after compensation

When finding the maximum peaks of the Rabi oscillation curves in Fig. 3 (a), we can obtain the optimal duration $\tau$ in order to achieve the maximum $\pi$ pulse fidelity for each diameter ratio, which are recorded as $\tau_i = \gamma_i \cdot \tau_0$ ($i = 1, 2, 3$) for the normal situation, and $\tau'_i = \gamma'_i \cdot \tau_0$ for the better situation respectively, in which $\gamma_i = \{1.051, 1.207, 1.373\}$, $\gamma'_i = \{1.019, 1.057, 1.127\}$, and $\tau_0 = \pi/\Omega_{max0}$. By replacing $\Omega_{max}$ with $\gamma_i \cdot \Omega_{max0}$ in equation (7), and calculating the integrals as section 3.2 do, we can obtain the improved Rabi oscillation curves as shown in Fig. 5. Compared with Fig. 3, we can see that a higher Raman $\pi$ pulse fidelity has been achieved with the same Raman pulse duration $\tau_0$. We also investigate the dependences of the $\pi$ pulse fidelity $F_\pi$ on

Raman pulse irradiation moment *t* in both situations with and without compensating Raman laser intensity. As shown in Fig. 6, though the π pulse fidelity still decreases with increasing free flight time *t*, compared with the corresponding initial value, the π pulse fidelity in the normal situation (red circles) has been improved by about 10% (pink rectangles) within the atom free flight time range of 0.4-1.6 s using this intensity compensation method. Similarly, the fidelity in the better situation (blue diamonds) has been improved by more than 10% (yellow triangles) within the atom free flight time range of 0.8-3 s.

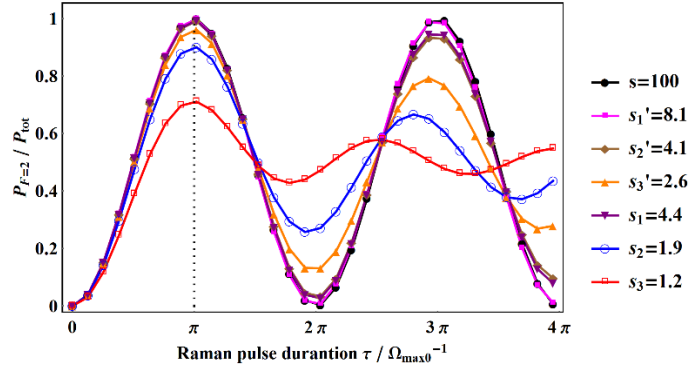

Fig. 5. Rabi oscillation curves after compensating Raman laser intensity. Compared with Fig. 3, a higher Raman π pulse fidelity has been achieved with the same Raman duration $\tau_0$.

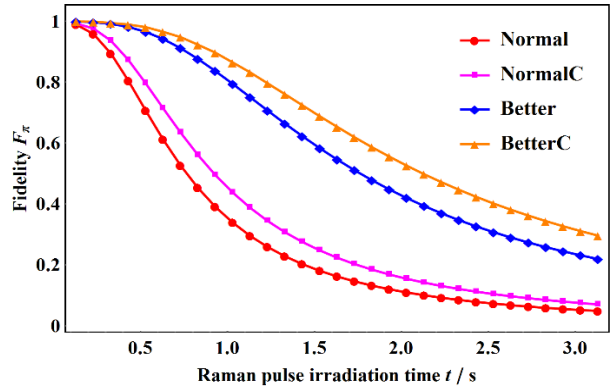

Fig.6. The dependences of the π pulse fidelity $F_\pi$ on Raman pulse irradiation time *t* in the normal (red circles) and better (blue diamonds) situations, and in the normal (pink rectangles) and better (yellow triangles) situations after compensating Raman laser intensity.

### 4.3 Fringe contrast after compensation

By replacing $\Omega_{max}$ with $\gamma_i \cdot \Omega_{max0}$ for the *i*th Raman pulse in equation (8) and calculating the integrals as section 3.3 do, we can obtain the interferometer fringes after compensation as shown in Fig. 7. Compared with the corresponding initial fringe, the fringe contrast has been improved by 13.6% from 41.5% (red circles) to 55.1% (pink rectangles) in the normal situation, and improved by 5.4% from 87.4% (blue diamonds) to 92.8% (yellow triangles) in the better situation. The dependences of fringe contrast on Raman pulse interval time $T$ are shown in Fig. 8. Though the fringe contrast still decreases with increasing Raman pulse interval time $T$, compared with the corresponding initial value, the contrast in the normal situation (red circles) has been improved by about 10% (pink rectangles) within the Raman pulse time interval range of 0-0.4 s, and the contrast in the better situation (blue diamonds) has been improved by more than 10% (yellow triangles) within the Raman pulse time interval range of 0.4-1.1 s.

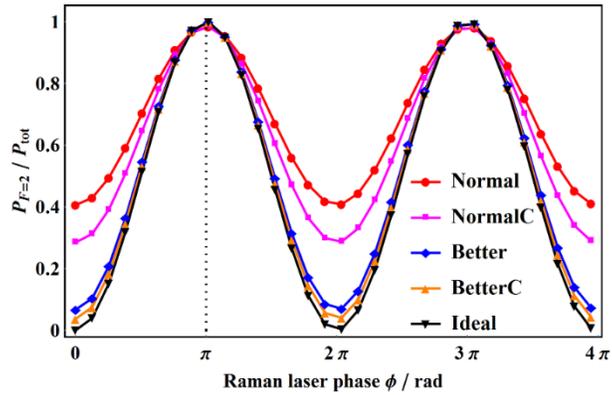

Fig. 7. Interferometer fringes before and after compensating Raman laser intensity. Using this intensity compensation method, the fringe contrast has been improved from 41.5% (red circles) to 55.1% (pink rectangles) in the normal situation, and improved from 87.4% (blue diamonds) to 92.8% (yellow triangles) in the better situation.

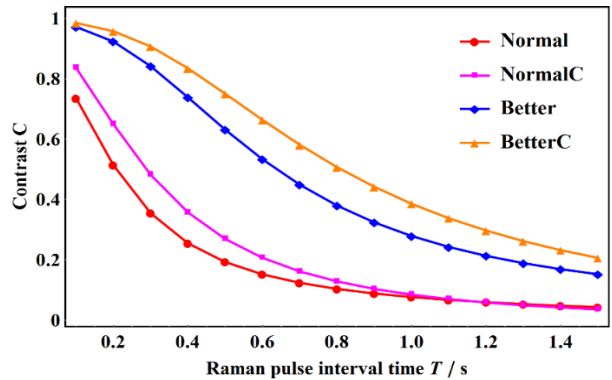

Fig. 8. Interferometer fringe contrasts as functions of Raman pulse interval time $T$ with and without compensating Raman laser intensity. Using this intensity compensation method, the contrast in the normal situation (red circles) can be improved by about 10% (pink rectangles) within the time interval range of 0-0.4 s, and the contrast in the better situation (blue diamonds) can be improved by more than 10% within the time interval range of 0.4-1.1 s.

## 5. Discussion

As mentioned above, the fidelity of the Raman π pulse and the fringe contrast are determined by the diameter ratio of Raman beam to atomic cloud, which means that we also can compensate the contrast reduction due to atom expansion if we could use a larger Raman beam size. However, a larger beam size requires a higher Raman laser power to maintain the same Rabi frequency, which namely requires expensive high power laser system. In practice, this is not an economic alternative. Besides, a larger Raman beam size requires a larger laser-beam-expanding and collimating telescope system, a larger mirror and λ/4 wave plate with good performance accordingly. These requirements are not easy to achieve in practice and might increase the wavefront aberrations related errors[35, 36]. Alternatively, we also can achieve the maximal π pulse fidelity by using a longer Raman duration as above. However, a longer Raman duration would results in a more strictly velocity selection effect[27] which, in return, would decay the fringe contrast more seriously. Therefore, our intensity compensation strategy is superior to the method of changing for a lager Raman beam system or using a longer Raman pulse duration.

Our theoretical simulation results demonstrate the general effects of the Raman laser intensity compensation for suppressing atom expansion induced contrast loss in atom interferometers. In these simulations, we integrate the atoms of the whole space range. In the experiments of [25, 35] and [22, 36], which corresponding to the normal and better experimental conditions respectively, the horizontal detection range are 20 mm and 13 mm respectively. These detection ranges are close to the atom diameters of 20.4 mm and 13.5 mm at the detection moment (0.77s after launch). However, in some experiments[38], the detection zone might be smaller than the atomic cloud size, which means that the atoms that are far away from the atomic cloud center are excluded from the detection signal. In this case, the integrating region should be modified according to the detection zone size and the improvement effects of this mitigation strategy might be decreased by a certain extent.

From the sensitivity function $S \propto 1/(k_{eff}T^2C\sqrt{N})$, it is obvious that using this mitigation strategy, the measurement sensitivity of the atom interferometer can be increased by the same ratio as contrast. In practical experiments, the improvement of the fringe contrast enable us to run the atom interferometer with a longer Raman pulse interval time $T$. Therefore, we also can run the atom interferometer with a visible fringe contrast and a larger available Raman pulse interval time

*T*. In this case, the measurement sensitivity can be increased quadratically.

## 6. Conclusion

In summary, we have investigated the mechanism of atom expansion related contrast reduction in atom interferometers. Simulation results show that the decrease of the diameter ratio of Raman beam to atomic cloud leads to a decrease of the $\pi$ pulse fidelity, a change of the $\pi$ pulse duration, and a significant reduction of fringe contrast. We proposed a mitigation strategy of increasing the intensities of the second and third Raman pulses, and evaluated the effects of this mitigation strategy in typical atom interferometer gravimeter conditions and in different Raman pulse interval conditions. Research results show that the fringe contrast can be improved by 13.6% using this Raman laser intensity compensation method. We also compared this strategy with other alternative strategies, discussed the possible correction when applying this theoretical model into other situations, and analyzed the improvement for measurement sensitivity. This mitigation strategy has potential applications in increasing the sensitivity of atom interferometer-based precision measuring, including precision measuring of the gravity, gravity gradient, rotation, and magnetic field gradient, as well as testing of the Einstein equivalence principle.

**Acknowledgments**

We thank Dr. Christian Freier, Bastian Leykauf, Dr. Vladimir Schkolnik, Dr. Markus Krutzik and Prof. Achim Peters from Humboldt University for enlightening discussions and a critical reading of the manuscript. This work is supported by the National Natural Science Foundation of China under Grant No. 51275523, Specialized Research Fund for the Doctoral Program of Higher Education of China under Grant No. 20134307110009, and the Graduate Innovative Research Fund of Hunan Province under Grant No. CX2014A002.